\documentclass[twocolumn,showpacs,amsmath,amssymb,prb,graphics,graphicx]
{revtex4}
 \usepackage{graphicx}
\usepackage{bm}
\usepackage{dcolumn}
\begin{document}

\title{On the field dependence of the vortex core size}

\author{V. G. Kogan and N. V. Zhelezina}
\affiliation{Ames Laboratory - DOE and Department of  Physics and Astronomy,
Iowa State University, Ames IA  50011-3160}

\begin{abstract}
We argue that in clean high-$\kappa$ type II superconductors, the
low temperature vortex core size (defined as   the
coherence length $\xi$) in high fields should decrease with
increasing applied field   in  qualitative agreement with
experimental data.  Calculations are done for the Fermi sphere and
cylinder (with the field parallel to the cylinder axis). The results
for clean materials at  $T=0$ can be represented as 
$\xi(H)/\xi(H_{c2}) = U ( H/H_{c2})$ with $U$ being an universal
function.  
\end{abstract}
\maketitle

\section{Introduction}
 
The coherence length $\xi$ has first been introduced as a
phenomenological length scale in the near-$T_c$ Ginzburg-Landau (GL)
description where it sets, among other things, the ``vortex core size"
$\rho_c$. Since the core, in fact, does not have a sharp boundary,
the size $\rho_c $ cannot be unambiguously defined and is commonly
chosen  ``operationally convenient", i.e., in a way which varies
from one experimental or theoretical situation to another. For
example,
$\rho_c$ may be defined as the radius of a circle where the
persistent current is maximum, \cite{Sonier1} or - within the London
approach - as the distance at which the divergent London current
density reaches the depairing value, or assuming the core being a
normal cylinder and equating the core contribution to the vortex
energy to $(H_c^2/8\pi)\pi\rho_c^2$ with $H_c^2/8\pi$ being the
condensation energy.\cite{deGennes,Tinkham} Another example comes
from the scanning tunneling work where ``the vortex core radius is
arbitrary defined by that distance $\rho_c$ from the vortex center,
for which the tunneling current has decreased from $I_{max}$ to 36\%
of
$I_{max}-I_{min}$."\cite{Golubov} Clearly, these procedures yield
different values of
$\rho_c$, although  all of them have the same order of magnitude.  

A lot of experimental effort has been invested   recently 
in study of the vortex core size, see  the 
review by Sonier and references therein.\cite{Sonier1} Notably,
whatever the definition of $\rho_c$ is adopted, the low temperature
 $\rho_c$ is shown to decrease with increasing field in a
number of materials such as NbSe$_2$, V$_3$Si, LuNi$_2$B$_2$C,
YBa$_2$Cu$_3$O$_{7-\delta}$, CeRu$_2$ physical characteristics
of which have little to do with each other (except all of them have 
a large GL parameter  $\kappa=\lambda_L/\xi$). The dependences
$\rho_c(H)$ for all tested materials are qualitatively similar; for
large fields $\rho_c\sim 1/\sqrt{H}$. Properties of the 
quasiparticle spectrum inside and outside the cores (where the
excitations may form narrow conducting bands) are considered 
responsible for the $H$ and $T$ dependences of $\rho_c$. 

The point of this paper is to provide a   
general theoretical  argument based on the weak-coupling BCS
theory in addressing causes of the field dependence of $\rho_c$. This
argument is omitted in the current discussion of the problem in the
experimental community, although a few examples of numerical
solutions of the microscopic equations of superconductivity show
that the $H$ dependence of $\rho_c$ follows from the  
theory under very general assumptions.\cite{Machida1,Machida2,Pedja} 
Our approach can be applied to the problem of $\rho_c(H)$ only in
large fields of high-$\kappa$ materials; still, we obtain our main
results analytically that has certain advantages as compared to 
however powerful numerical techniques and enables one to make
experimentally verifiable predictions. \\

We begin with the notion that within the microscopic
theory, at arbitrary magnetic fields $H$ and temperatures $T$, it
is not clear what exact value one should assign to  
$\xi$. The difficulty comes from the fact that $\xi$ and $\rho_c$ are 
not among the basic input parameters of the theory; instead, they 
should be calculated, analytically a very difficult if at all 
 possible task in general.  There is, however, a region at the second
order phase transition from  superconducting to  normal state,
the SN boundary, where the theory can be linearized and consequently
$\xi$ is well defined.  

The linearization has been performed in the seminal work of Helfand
and Werthamer (HW) on the upper critical field $H_{c2}(T,\tau)$ with
$\tau$ being the mean scattering time on nonmagnetic
impurities.\cite{HW}  They have shown that at $H_{c2}$ where  the
order parameter $\Delta$ goes to zero, it satisfies for any $T$ a
linear  equation,
\begin{equation}
-\xi^2(T)\,\Pi^2\Delta=\Delta\,,\label{linearGL}
\end{equation}
where $\bm\Pi=\nabla+2\pi i\bm A/\phi_0$ is the gauge invariant
gradient, $\bm A$ is the vector potential, and $\phi_0$ is the flux
quantum. Formally, the equation is equivalent to the
Schr\"odinger equation for a charge in uniform magnetic field; the
field $H_{c2}$ at which superconductivity first nucleates corresponds
to the lowest eigenvalue of this equation: $H_{c2}=\phi_0/2\pi\xi^2$.
This field (and $\xi$) is obtained by solving the basic
self-consistency equation of the theory:
\begin{equation}
\frac{\hbar}{2\pi T}\,\ln\frac{T_c}{T}
=\sum_{\omega=0}^{\infty}\left(\frac{1}{\omega}-\frac{2\tau
S}{\beta-S}\right)\,,
\label{self*}
\end{equation}
   where the function $S(T,\xi,\tau)$ can be written as:
\begin{eqnarray}
S&=&\frac{2\beta}{\ell
q}\int_0^{\infty}ds\,e^{-s^2}\tan^{-1}\frac{s\ell
q}{\beta}\label{HWint}\\ 
&=&\sum_{j=0}^{\infty}\frac{ 
j!}{2j+1}\,\left(-\frac{\ell^2q^2}{\beta^2}\right)^j\,,\label{HWsum}\\
\beta&=&1+2\omega\tau\,,\quad q^2=2\pi H_{c2}/\phi_0=\xi^{-2}\,.
\end{eqnarray}
Here, $\omega=\pi T(2 n+1)/\hbar$, $n$ is an integer, $v$ is the
Fermi velocity,  and
$\ell=v\tau$ is the mean-free path. The power series representation
of $S$ is obtained by formally expanding tan$^{-1}$ and then
integrating over $s$.  The evaluation of $S$ was performed for the
isotropic Fermi surface, i.e., for a Fermi sphere. 

Thus, strictly speaking, the length $\xi$ is defined only at the
SN phase boundary $H_{c2}(T)$, and the question remains whether or not
the same definition of $\xi$ is useful out of the immediate vicinity
of $H_{c2}(T)$. In fact, in a variety of situations (small 
samples, proximity systems) the SN  transition may
take place far from the bulk $H_{c2}(T)$. To approach the problem of
the phase boundary in these systems, one has to know $\xi(H,T)$ in
a broad domain of the $H-T$ plane away of the bulk $H_{c2}(T)$.  

A method to evaluate $\xi(H,T)$ had been developed in 
Refs.\,\,\onlinecite{K1} and \onlinecite{K2}. In principle, the method
follows   HW by utilizing   the field {\it uniformity} and the 
$\Delta$ {\it smallness} at the SN transition wherever it occurs. 
Below, we outline the method as applied to the three-dimensional
(3D) isotropic case of a Fermi sphere. Then, we consider 
  2D isotropic materials, i.e., the Fermi cylinder.  We find 
that $\xi(H,T)$ so obtained {\it decreases} when $H$
{\it increases} toward $H_{c2}$ away of the critical temperature $T_c$
in clean superconductors; the effect is suppressed by impurity
scattering and is absent in the dirty limit. We provide a closed form
equation  for $\xi(H)$ for zero-$T$ clean case for both the Fermi
sphere and Fermi cylinder, and show that the results can be
represented as 
\begin{equation}
\frac{\xi(H)}{\xi(H_{c2})} = U\left(\frac{H}{H_{c2}}\right) 
\label{univ}
\end{equation}
with $U$ being an universal function. 

We next argue that the same procedure can be applied to the mixed
state in applied fields $H<H_{c2}$ near the vortex core centers in
materials with large   
$\kappa =\lambda_L/\xi$. This is because near the centers the
field (varying  on the scale of the London penetration depth
$\lambda_L$) can be taken as {\it uniform} and $\Delta(r)\to 0$   at
the center. We find our results 
 in qualitative agreement with the data available,  
uncertainties of experimental procedures of extracting $\xi$
notwithstanding.

\section{Fermi sphere: $ \bm {H_{c2}(T)}$   and $\bm
{S(H,\xi,\omega,\ell)}$}

Here we reproduce major points of the $\xi(H)$ derivation for the
system near the second order SN transition with the help of the
quasiclassical Eilenberger formalism.\cite{E} The main equations of
the theory read:
\begin{eqnarray}
&&\tau\bm{v\cdot\Pi}f=g(F+2\tau\Delta)-(G+2\omega\tau)f\,,\label{Eil1}\\
&&\frac{\Delta}{2\pi T}\,\ln\frac{T_c}{T}
=\sum_{\omega=0}^{\infty}\left(\frac{\Delta}{\hbar\omega}-F\right)\,.
\label{Eil2}
\end{eqnarray}
Here, $\bm v$ is the Fermi velocity; $f(\bm r,\omega,\bm v)$ and
$g(\bm r,\omega,\bm v)$ are the Eilenberger Green's functions with
averages over the Fermi surface denoted as $F=\langle f\rangle$ and
$G=\langle g\rangle$.  

In the normal phase $f=0$ and $g=1$. In a
small vicinity of the SN transition, $|f|\ll 1$, whereas $g$ can
still be set unity in linear approximation in $f$ due to
normalization $g=(1-ff^\dagger)^{1/2}$ (for the same reason we do
not need here an equation for $f^\dagger$). Equation (\ref{Eil1})
can be linearized:
\begin{eqnarray}
&&   \bm{\ell\cdot\Pi}f = {\tilde F}-\beta f\,,\label{linear}\\
&& \bm \ell=\bm v\tau\,,\quad {\tilde F} =F+2\tau\Delta/\hbar\,,\quad
\beta=1+2\omega\tau.
\label{ell}
\end{eqnarray}
The solution of Eq.\,(\ref{linear}) is  written as
\begin{equation}
f=(\beta+\bm{\ell\cdot\Pi})^{-1}{\tilde F}=\int_0^{\infty}d\rho\,
e^{-\rho(\beta+\bm{\ell\cdot\Pi})}{\tilde F}\,,
\end{equation}
or for the Fermi surface average:
\begin{equation}
F= \int_0^{\infty}d\rho\,e^{-\rho\beta}\langle
e^{-\rho \bm{\ell\cdot\Pi}}{\tilde F}\rangle\,. \label{F}
\end{equation}
We now assume that $\Delta, F,$ and ${\tilde F}$ satisfy
Eq.\,(\ref{linearGL}); then, utilizing  commutators  of
the operator 
$\bm \Pi$ in an {\it uniform} field and the known properties
of exponential operators,\cite{K1} one can manipulate Eq.\,(\ref{F})
to
\begin{equation}
F(\bm r,\omega)=\Delta(\bm r)\,\frac{2\tau S}{\beta-S}\,,\label{FF}
\end{equation}
where
\begin{eqnarray}
S&=&\sum_{m,j=0}^{\infty}\frac{(-q^2)^j 
}{j!(2m+2j+1)}\left( \frac{(m+j)!}{m!}\right)^2\left(
\frac{\ell^2 }{\beta^2}\right)^{m+j}\nonumber\\
&\times& \prod_{i=1}^m [(2i-1)q^2-\xi^{-2}]\,,\qquad
q^2=\frac{2\pi H}{\phi_0}.\label{s-sum}
\end{eqnarray}
After substituting $F$ of Eq.\,(\ref{FF}) in the self-consistency
Eq.\,(\ref{Eil2}) and cancelling   $\Delta(\bm r)$, we
obtain an implicit Eq.\,(\ref{self*}) for $\xi(H,T,\ell)$. It is
easy to see that at $H_{c2}$ where $q^2=1/\xi^2$, the series
(\ref{s-sum}) reduces to the HW sum (\ref{HWsum}). 

The double sum (\ref{s-sum}) is in fact an asymptotic series and is
difficult to deal with when the goal is to solve Eq.\,(\ref{self*}) 
for $\xi(H,T,\ell)$. The situation simplifies greatly in the dirty
limit: $S$ is an expansion in powers of $\ell$. Keeping only the
terms with $m+j=0,1$, one obtains
\begin{equation}
S=1-\frac{\ell^2}{3\xi^2\beta^2}\,,\quad \ell^4/\xi^4\ll 1\,,\quad
\ell^4q^4\ll 1\,.\label{trancS}
\end{equation}
When substituted in the self-consistency Eq.\,(\ref{self*}), this
yields de Gennes-Maki dirty limit result for $\xi(T,\ell)$ and
$H_{c2}(T)$.\cite{deGennes} Since  the field $q^2$ does not
enter $S$,   {\it the coherence length in the dirty limit is
field independent}. In other words, in the dirty limit, the coherence
length  at a given $T$ determined at the upper critical field, is the
same at this $T$ for any $H$.

Formally similar situation takes place near the critical temperature
$T_c$ where the truncation (\ref{trancS}) is justified by smallness
of  $q^2$ and $\xi^{-2}$ (not of $\ell$). We conclude that {\it
near $T_c$ the coherence length is field independent for any $\ell$}.
We then expect the strongest $H$ dependence of $\xi$ to exist at low
temperatures in clean materials. 

Given the complexity of series (\ref{s-sum}), it is desirable to
have an integral representation for $S$ better suited for analytic
and numerical work. This had been done in Ref.\,
\onlinecite{K2} for the 3D case of a spherical Fermi surface. We
refer the reader for details of this nontrivial procedure  and
provide here the result:
\begin{eqnarray}
&&S(u,\sigma)=\sqrt{\pi}\,{\rm
Re}\int_0^{\infty}ds\frac{(1-us^2)^{\sigma-1}}{(1+us^2)^{\sigma}}\,{\rm
erfc}\,s\,,\label{S3D}\\
&&u=\frac{\ell^2q^2}{\beta^2}\,,\qquad
\sigma=\frac{1}{2}\left(1+\frac{1}{\xi^2q^2}\right)\,;\label{u}
\end{eqnarray}
 ${\rm erfc}\,s=(2/\sqrt{\pi})\int_s^{\infty}dz\exp(-z^2)$ ($\sigma$
here differs by the sign from that used in Ref.\,\onlinecite{K2}).
Note that $\sigma=1$ at $H_{c2}$; integration by parts in
(\ref{S3D}) gives the HW integral (\ref{HWint}). One can check by
formally expanding the integrand in powers of $us^2$ and integrating
over $s$, that the integral (\ref{S3D}) can indeed be written as the
series (\ref{s-sum}), see Appendix B in Ref.
\onlinecite{K2}.

\section{Fermi cylinder}

The calculations of the previous section cannot  be done
for an arbitrary Fermi surface. Still, for some simple shapes it is
possible. The simplest of those is the Fermi cylinder with the field
parallel to the cylinder axis. Employing the same procedure  
outlined for the Fermi sphere, we arrive at
\begin{equation}
S=\sum_{m,n=0}^{\infty}\frac{(-1)^n2^m 
(2m+2n)!}{n!(m!)^2}\,\left(\frac{u}{4}\right)^{m+n}(1-\sigma)_m\,,
\label{S2Dsum} 
\end{equation}
where $u$ and $\sigma$ are defined in Eq.\,(\ref{u}); we use the 
  notation 
$(1-\sigma)_m=(1-\sigma)(2-\sigma)\cdot\cdot\cdot(m-\sigma)$.\cite{Abr}
The integral representation of this sum can be obtained in a manner 
similar to that described in Ref. \onlinecite{K2} for the 3D case:
\begin{eqnarray}
S(u,\sigma)=\frac{2}{\sqrt{\pi}}\,{\rm
Re}\int_0^{\infty}ds\,\frac{(1-us^2)^{\sigma-1}}{(1+us^2)^{\sigma}}\,e^{-s^2}
\,;\label{S2D}
\end{eqnarray}
see Appendix A. To verify this result we write:
\begin{eqnarray}
\frac{(1-s^2u)^{\sigma-1}}{(1+s^2u)^{\sigma }}=\Big(\frac{1-s^2u }{
1+s^2u }\Big)^{ \sigma-1}\frac{1}{1+s^2u}\nonumber\\
=\sum_m\Big(\begin{array}{c}\sigma-1\\m\end{array}\Big)\Big(
-\frac{2us^2}{1+us^2}\Big)^m\frac{1}{1+s^2u}\nonumber\\
=
\sum_{m,n}\frac{(-1)^\mu\mu !(\sigma-1)_m}{(m!)^2n!}2^mu^{\mu}
s^{2\mu}\,, 
\end{eqnarray}
where   $\mu=m+n$. 
Substituting this to Eq. (\ref{S2D}) and integrating over $s$ one
obtains $S$ in the form (\ref{S2Dsum}). 

 Having the quantity $S(H,\xi,T,\ell)$ for both 3D (Fermi sphere) and
2D (Fermi cylinder),  one can solve Eq.\,(\ref{self*}) for 
$\xi(H,T,\ell)$  that in general can be done numerically. As
pointed out above, the most interesting is the situation at $T=0$ in
clean materials; this case can be treated analytically.

\section{Clean materials at $\bm {T=0}$}

To deal with the divergence of $\ln(T_c/T)$ in Eq.\,(\ref{self*})
we  note than the sum over $\omega$ on the right-hand side
(RHS) is actually extended to the Debye frequency $\omega_D$.
Then, we have for the finite sum  
\begin{equation}
 \sum_{\omega >0}^{\omega_D} 
\frac{1}{\hbar\omega} \approx\frac{1}{2\pi
T}\,\ln\frac{2\hbar\omega_D e^{\gamma}}{\pi T},
\label{e18}
\end{equation}
where the neglected terms are of the order $T^2/\hbar^2\omega_D^2$
and
$\gamma\approx 0.577$ is the Euler constant. Hence the divergent
$\ln T$ in Eq.\,(\ref{self*}) drops off. Since  in the
clean limit $\beta\approx 2\omega\tau$ and 
$2\tau S/(\beta-S)\approx S/\omega$ we obtain instead of
Eq.\,(\ref{self*}) in the zero-$T$ limit: 
\begin{equation}
 \ln\frac{2\hbar\omega_D}{\Delta_0}=2\pi T\sum_{\omega
>0}^{\omega_D} \frac{S}{\hbar\omega}
\rightarrow\int_0^{ \omega_D}\frac{d\omega}{\omega}\,S(u,\sigma)\,.
\label{e19}
\end{equation}
 The integral at the RHS diverges logarithmically for
$\omega_D\to\infty$, and so does the LHS; in other words, $\omega_D$
should drop off the result.  This integral is evaluated  in Appendix
B for both 3D and 2D cases. 

We then obtain an implicit equation for $\xi$:  
\begin{eqnarray}
\ln\frac{\hbar vq }{\Delta_0\alpha} +\frac{\cos(\pi\sigma)}{4}
\left[\psi\Big(\frac{1+\sigma}{2}\Big)-
\psi\Big(\frac{\sigma}{2}\Big)\right]+\frac{\psi(\sigma)}{2}
=0,\nonumber\\ 
\, \label{***}
\end{eqnarray}
  where $\psi$ is the Digamma function. The only difference between 2D
and 3D situations is in the number  $\alpha$:
\begin{equation}
\alpha_{2D}= \sqrt{2}\,,\qquad \alpha_{3D}=e/ \sqrt{2}\,. 
\label{alphas}
\end{equation}

Setting $\sigma=1$ in Eq.\,(\ref{***}) one obtains:
\begin{equation}
H_{c2}=\frac{\phi_0}{2\pi\xi_{c2}^2}\,,\qquad 
\xi_{c2}=\frac{\hbar v }{\Delta_0\sqrt{2}\alpha}\,e^{-\gamma/2}\,.
\label{xi_c2*}
\end{equation}
This yields for the 3D case:
\begin{equation}
H_{c2}(0)=\frac{\phi_0\Delta_0^2}{2\pi\hbar^2v^2}\,
e^{2+\gamma} \,,
\label{Hc2(0)}
\end{equation}
the value obtained variationally by Gor'kov\cite{gorkov}  
and proven to be exact by HW; in HW reduced units it corresponds to
$h^*(0)=H_{c2}(0)/T_cH_{c2}^{\prime}(T_c)\approx 0.72$.\cite{HW} For
the 2D case, this gives $h^*(0) \approx 0.59$, the result   
obtained by Bulaevskii.\cite{Lev}
 
We now observe that material parameters enter Eq.\,(\ref{***})
only in the first term under the log-sign. If one measures the
length in units of $\xi_{c2}$ and uses the reduced field  
$h=H/H_{c2}$,  Eq.\,(\ref{***}) takes the form independent of
material parameters:
\begin{eqnarray} 
&&\ln(2h e^{\gamma})+\frac{\cos(\pi\sigma)}{2}
\left[\psi\Big(\frac{1+\sigma}{2}\Big)-
\psi\Big(\frac{\sigma}{2}\Big)\right]+ \psi(\sigma) 
=0,\nonumber\\ 
&&\sigma=\frac{1}{2}\left(\frac{1}{h\,\xi^{*2}}+1\right)\,,\qquad
\xi^*=\frac{\xi}{\xi_{c2}}\,.
\label{universal***}
\end{eqnarray}
Hence, this equation defines an universal curve $\xi^*(h)$
independent of either material characteristics $v_F,\Delta_0$ or
the dimensionality.  Given this curve and 
$H_{c2}(0)$, one can recover $\xi(H)$ for a clean
material at $T=0$.  
 
\begin{figure}[htb]
   \vspace{0.1in}
  \centering
  \includegraphics[width=80mm]{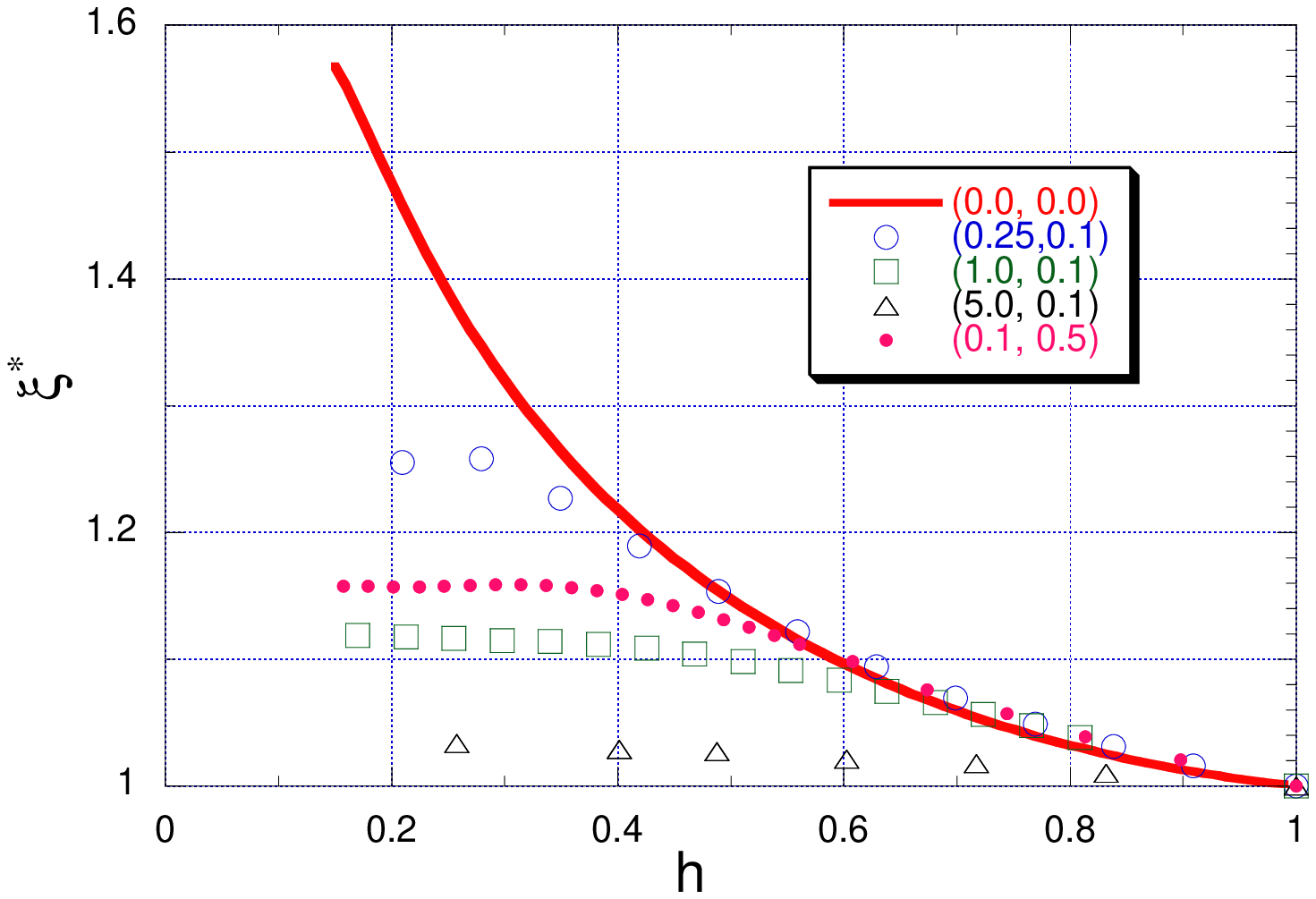}
\includegraphics[width=80mm]{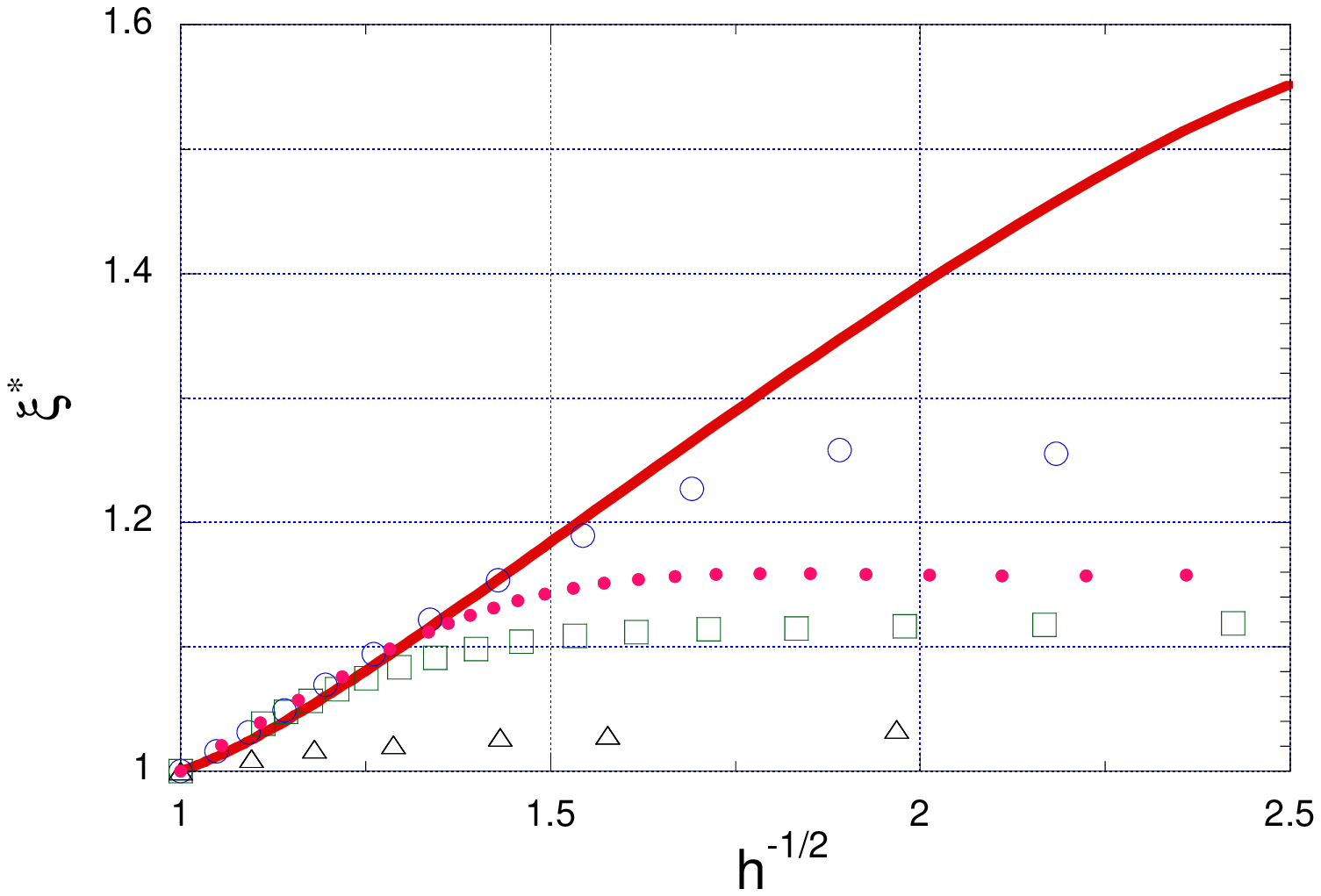}
    \caption{(Color online) The upper panel: the normalized
coherence  length
$\xi^*=\xi(H)/\xi(H_{c2})$   versus $h=H/H_{c2}$. The solid
line is calculated with the help of Eq.\,(\ref{universal***}) for
the clean limit at $T=0$. The open simbols show $\xi^*(\lambda,t,h)$
for a few values of the scattering parameter $\lambda=\hbar v/2\pi
T_c\ell$ and reduced temperatures $t=0.1$ shown in pairs $
(\lambda,t) $ on the legend. The full symbols are for a clean
material at an elevated temperature: $ (\lambda,t) =(0.1,0.5)$. The
lower panel:   $\xi^* $ versus $1/\sqrt{h}$. 
 }
 \label{fig1}
  \end{figure}

The curve $\xi(H)/\xi(H_{c2}) = U ( H/H_{c2})$ is shown as a solid
line in Fig.\,{\ref{fig1}} for $0.15<H/H_{c2}<1$; the reason why the
small fields domain is not shown is given in the next section. Also
shown are results of numerical evaluation of $\xi(H)$ for a few
values of the impurity parameter $\lambda=\hbar v/2\pi T_c \ell$. The
numerical calculation is done with the help of the self-consistency
Eq.\,(\ref{self*}) for arbitrary $T$ and $\lambda$;
$S(\xi,q^2,T,\lambda)$ is evaluated using an explicitly real form
given in Appendix C. 

It is worth noting that effect of raising temperature on $\xi(H)$ is
qualitatively similar to that of the impurity scattering, see solid
dots for $t=T/T_c=0.5$: both suppress the field dependence of $\xi$.
However, at low temperatures for reasonably clean materials in a 
broad  domain of high fields, $\xi(H)$ is well
represented by the zero-$T$ clean-limit curve; it is seen from the
upper panel of Fig.\,\ref{fig1} that for    
$\lambda=0.25$ this domain extends down to $h\approx 0.4$. 

The lower panel of Fig.\,\ref{fig1} shows the same results plotted
against $1/\sqrt{h}$, the quantity proportional to the intervortex
spacing. In this manner the data are often presented to examine 
possible connection between the field dependence of the
core size $\rho_c(H)$ and other properties of the mixed
state.\cite{Sonier1}  Our result shows that for materials on the
clean side with $\lambda<1$, the slope $d\xi^*/d(h^{-1/2})$ for $H\to
H_{c2}$ is universal.  In fact, using
Eq.\,(\ref{universal***})  this slope at $H_{c2}$ can be
evaluated for the clean limit at $T=0$ : 
\begin{equation}
\frac{d\xi^*}{d(h^{-1/2})}\Big|_{h=1}=1-\frac{8}{\pi^2}\approx
0.189\,.
\label{slope(1)}
\end{equation}
We also observe that for  real materials with $\lambda\ne 0$ and
$T\ne 0$,  $\xi^*(h)$ becomes flat as the field
decreases with the impurity and temperature dependent plateaus.

\section{The core size}

The above discussion of $\xi(H)$ applies at the SN second order phase
transition where the field is {\it uniform} and the  order
parameter $\Delta$ goes to zero. In fact, these conditions are met
in vortex cores of high-$\kappa$ type-II superconductors in high
fields. Indeed, in this case the field within the core of a size
$\xi$ is practically uniform since it varies on a much larger scale
$\lambda_L$. Besides, when one approaches the vortex center,
$\Delta\to 0$. To evaluate $\xi$ in this situation, one can use the
same formalism as at the SN phase boundary; in other words, the 
above procedure of evaluating $\xi(H)$ can be used to characterize
the  size $\rho_c$.  The core size so defined is, of course,  
``operationally convenient" from   our point of view just
as various definitions mentioned in the Introduction.
Our approach, however, is advantageous because along with  
evaluation of the $H$ dependence of the ``core size" $\xi$, we
predict that 
 \begin{itemize}
\item this dependence is weakened by scattering and
disappears in the dirty limit, 
\item  the $H$ dependence of  $\xi$
vanishes as $T\to T_c$, 
\item  $\xi(H)$ is  weakly affected by 
peculiarities of the Fermi surface, i.e., we expect qualitatively the
same dependence for various materials, 
\item  in reduced variables, the dimensionless 
coherence length $\xi^*=\xi/\xi_{c2}$ should be nearly universal
function of the reduced field $h= H/H_{c2}$ for clean materials in
high fields and low temperatures.  
 \end{itemize}
Therefore, our ``theoretically convenient" definition of the core size
can be checked experimentally. 

One should put a note of caution on our claim of ``universality" as
stated in the last two points. This feature expressed in
Eq.\,(\ref{universal***}), has been derived  for two simple
Fermi surfaces, a sphere and a cylinder.\cite{remark2} Nevertheless,
since the Fermi surface shape always enters   calculations of
macroscopic parameters as $\xi$ or $\lambda_L$ via averaging over
the whole surface, one does not expect the fine features of the
  surface to alter drastically our conclusion (except in 
special circumstances, e.g., when the local density of states has
sharp maxima at the surface). With this caveat we will use the term 
``universality" in further discussion.    

There is another drawback to our approach. When applied
to the SN phase boundary, say, of a proximity sandwich, the field
$H$ in the $\xi(H)$ dependence is the externally controlled
uniform applied field.  Defining the core size as $\xi(H)$, we imply
that $H$ is a the field value at the vortex center, $H_0$, which is
nearly constant  within the core provided $\kappa\gg 1$. However,
the problem is that there is no reliable and generally applicable
estimate of $H_0$, except numerical results with a particular choice
of parameters for low-$\kappa$ and for high-$T_c$   
materials.\cite{Machida1,Machida2,Pedja}  An exception is the case
of isolated vortices (the applied field $H_a\to 0$) in the GL domain,
where $H_0\approx 2H_{c1}$.\cite{Abrikosov} In a more interesting
situation of $H_a\gg H_{c1}$, the vortex fields are strongly
overlapped, and variations of the actual field within the vortex
lattice are weak relative to the applied field; in fact, they are on
the order of $H_{c1}\ll H_a$. Therefore, an error made by
considering the field at the vortex axes as equal to the applied
field is small. In other words, defining the vortex core size
$\rho_c(H_a)$   as $\xi(H_a)$   has a reasonable chance of success
only in large fields $H_a\gg H_{c1}$ and improves as $H_a\to
H_{c2}$.  
 
There is also a theoretical difficulty we encounter attempting   
to extend the analysis to $h\to 0$. In fact, the curve shown in
Fig.\,\ref{fig1} and generated by solving Eq.\,(\ref{universal***})
shows oscillating behavior if extended to low fields beyond
$h\approx 0.15$.\cite{remark1} Our numerical work for finite
$\lambda$ and $T$ shows that these oscillations are washed out
quickly with increasing  scattering and/or temperature and
are hardly seen for $\lambda > 0.25$, i.e., in still rather clean
materials.

\section{Comparison with data}

Relating the results obtained to  information available
on the vortex core size, we focus on the $\mu$SR data reviewed
recently by Sonier.\cite{Sonier1} This technique allows one to
obtain  the field  distribution  $h(\bm r)$ within the vortex
lattice. Then one can calculate the  current
distribution and define the core radius $\rho_c$ as the distance
from the vortex axis to the current maximum in the nearest neighbor
direction. This definition of the core size is independent of a
model one may choose to theoretically describe the distribution
$h(\bm r)$, the point stressed in Ref. \onlinecite{Sonier2}. We will
consider here the data on so defined $\rho_c$. 

 The $\mu$SR data on $h(\bm r)$ can also 
be  analyzed with the help of the London model or its nonlocal
version. For simplicity, we consider here the standard London
isotropic result:
\begin{equation}
h(\bm r) = B\sum_{\bm G}\frac{e^{i\bm G\cdot\bm
r}}{1+\lambda_L^2G^2}
\label{London}
\end{equation}
where $B$ is the magnetic induction and the sum is extended over the
reciprocal lattice $\bm G$.  

 It is relevant for this discussion that (a) the London model 
contains only one length scale, the penetration depth 
$\lambda_L$, and (b) the model implies  the constant order
parameter $\Delta$  and therefore breaks down at distances of the
order $\xi$. The latter comes about formally in
Eq.\,(\ref{London}) since  the sum   is divergent (this is readily
seen as the logarithmic divergence of $h$ when $r\to 0$).  To mend
this inherent shortcoming of the London model, various cutoffs are
commonly used,  e.g., by introducing a factor $\exp(-{\rm
const}\,G^2\xi^2)$ which excludes distances smaller than $\xi$.
Numerous efforts to fix the constant's value notwithstanding (see,
e.g., Ref.\,
\onlinecite{Brandt}), in practice this constant is used quite
liberally depending on the  application in question. Other
cutoffs basically suffer of similar uncertainties.\cite{Clem} Hence,
the {\it reliable} results of the London model are only those that
are insensitive to the cutoff chosen. Still, one can fit the data
$h(\bm r)$ to a properly truncated sum (\ref{London}), and extract
the best-fit parameters $\lambda_L$ and $\xi$ along with their $H$
dependence. Interestingly enough, the so extracted $\xi(H)$ behaves
as a function of $H$ in nearly the same manner as $\rho_c(H)$
extracted directly from the field distribution; it is found for a few
materials that in high fields $\rho_c\approx \xi+C$  with a material
dependent constant $C$.\cite{Sonier1}    

\begin{figure}[htb]
   \vspace{0.1in}
  \centering
  \includegraphics[width=80mm]{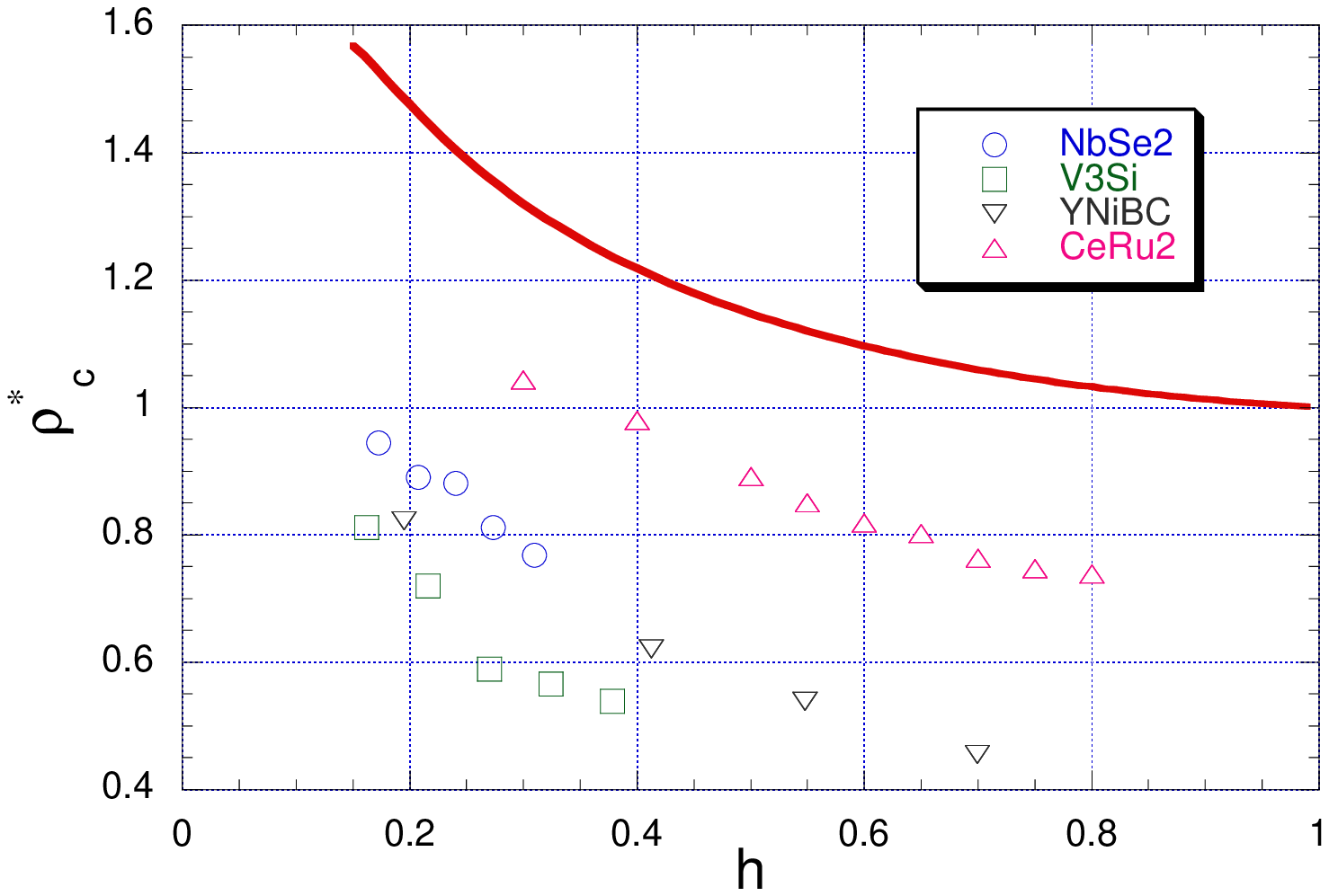}
 \includegraphics[width=80mm]{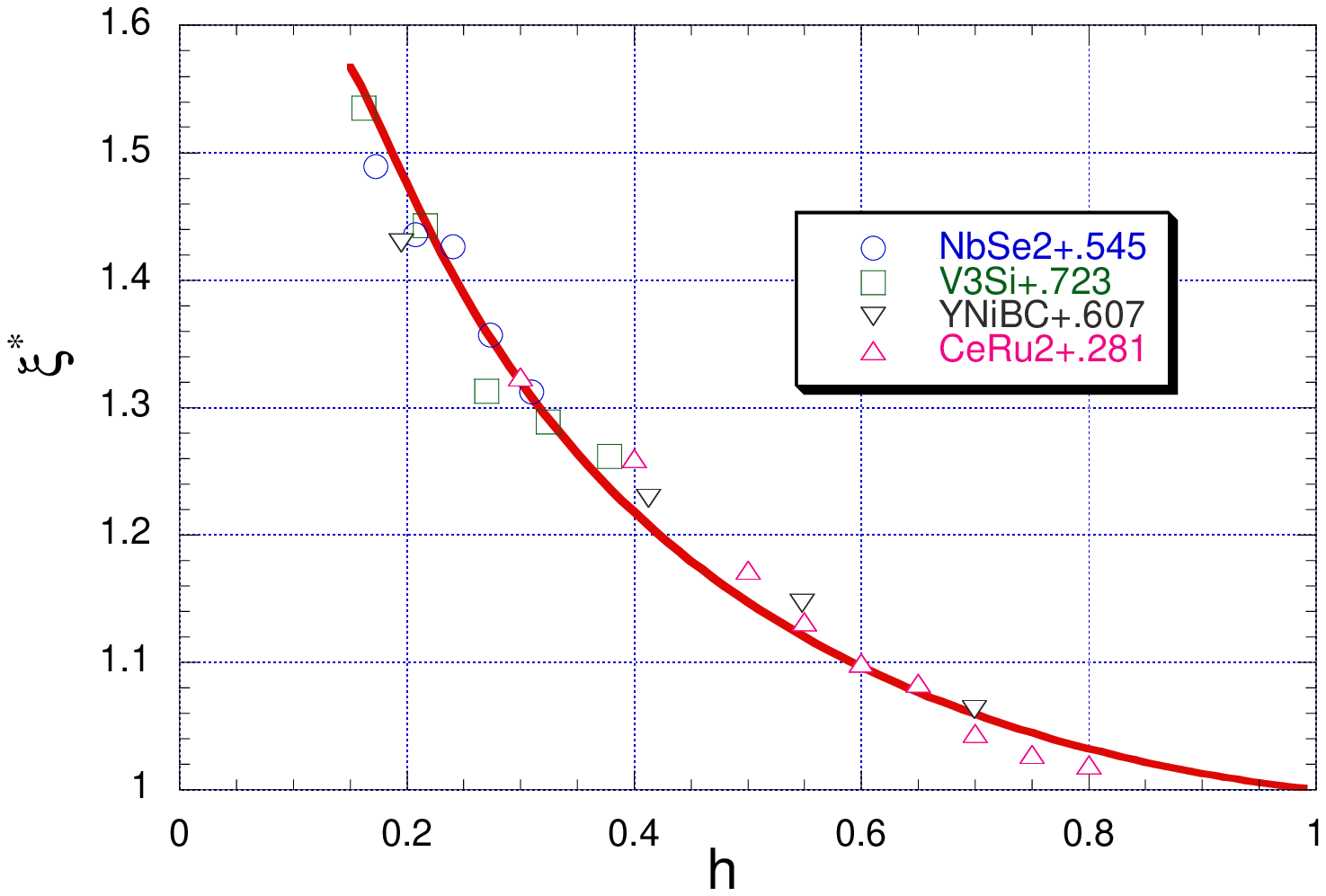}
 \includegraphics[width=80mm]{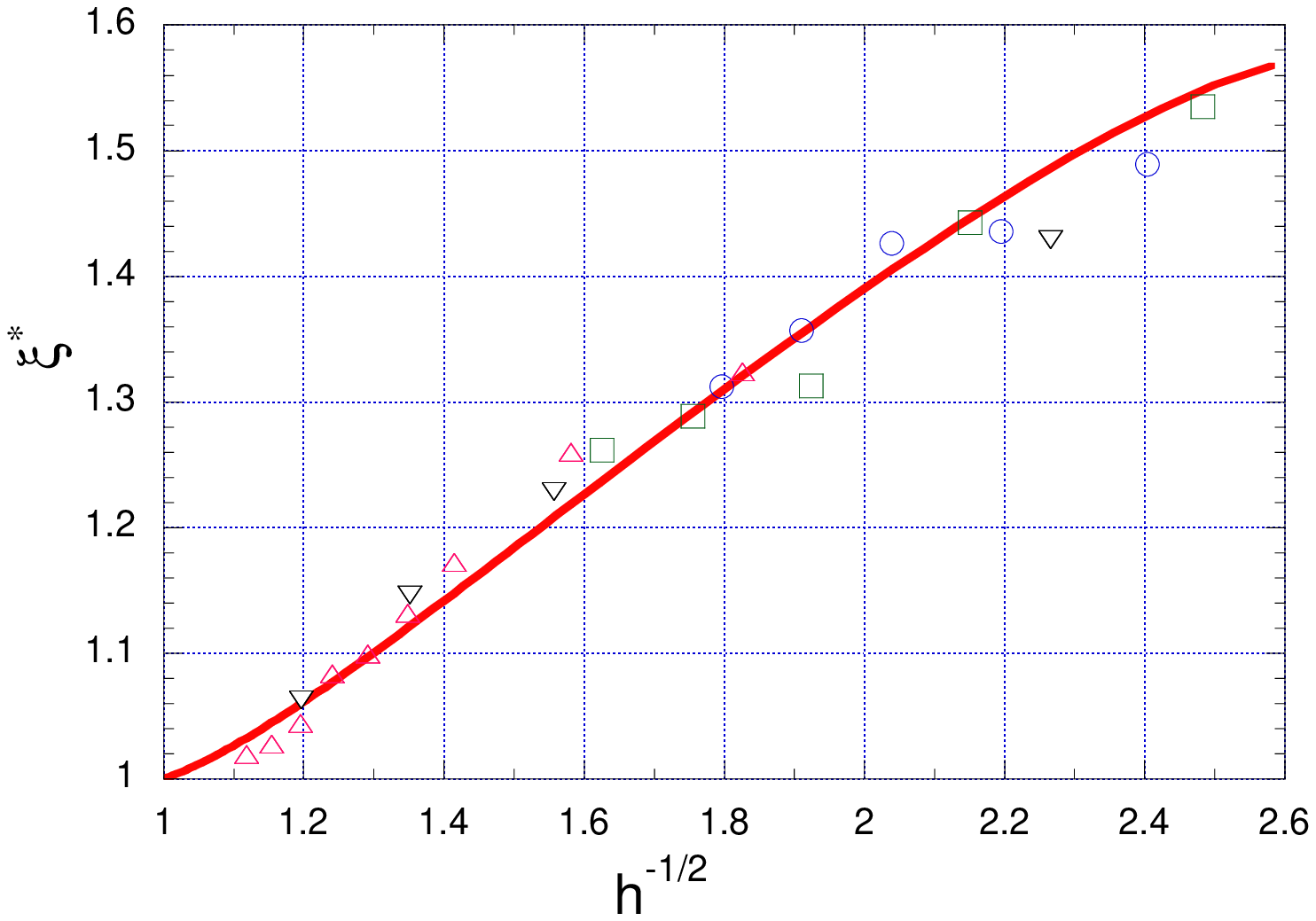}
    \caption{(Color online) The upper panel: the experimental core
radius $\rho_c$
 normalized on $\xi_{c2}=\sqrt{\phi_0/2\pi H_{c2}(T)}$ of each
compound for materials indicated in the legend; $T$ is the
temperature of each experiment. The solid line is the theoretical
$\xi^*(h)$ with $h=H/H_{c2}(T)$, the same curve as in
Fig.\,\ref{fig1}. The middle panel: the same data shifted by amounts
indicated in the legend. The lower panel: the same as the middle
panel, but plotted versus $1/\sqrt{h}.$
 }
 \label{fig2}
  \end{figure}

We now consider the data on $\rho_c(H)$ for  
V$_3$Si, NbSe$_2$, YNi$_2$B$_2$C, and CeRu$_2$ provided in
Refs.\,\onlinecite{NbSe2,Sonier2,YNiBC,CeRu2}, respectively, and
summed up in the review by Sonier.\cite{Sonier1} All the samples are
high quality single crystals and have large GL parameters $\kappa$;
we assume them clean   (the available scattering parameters are
$\lambda(V_3Si)\approx 0.13$ and $\lambda(NbSe_2)\approx 0.15$). The
reduced temperatures of the $\mu$SR experiments were low:  
$\approx$   0.22, 0.33, 0.19, and 0.3, respectively.  For each
material we have taken the $H_{c2}$ at a corresponding temperature,
  calculated $\xi_{c2}$, and normalized the experimental core
radius to this value to obtain $\rho_c^*=\rho_c/\xi_{c2}\,$. The
results are plotted in the upper panel of Fig.\,\ref{fig2} together
with the theoretical $\xi^*$ versus reduced fields $h=H/H_{c2}$. For
reasons explained above we took only the data points for $h>0.15$.
We expect the experimental $\rho_c^*(h)$
and the theoretical $\xi^*(h)$ to be  shifted by a
material, temperature and purity dependent constant:
$\xi^*(h)\approx \rho_c^*(h)+C(\lambda,T)$. Since the temperatures
and impurity parameters in different experiments were different,  we
do not expect these shifts to be the same for the materials
examined. In this situation, we have chosen the constants $C$ so as
to shift the data points as close as possible to our curve of
$\xi^*(h)$. The result is shown in the middle panel of
Fig.\,\ref{fig2}; the shifts needed are shown in the panel
legend. Although the shifts vary, the data for different
materials land nicely in a vicinity of our curve. This
supports our guess of ``universality", a considerable ambiguity
of the procedure notwithstanding. 

 An interesting feature of the data and of the universal curve
$\xi^*(1/\sqrt{h})$  is seen in the lower panel of Fig.\,\ref{fig2}:
 the slope of this curve starting with the value (\ref{slope(1)}) of
$\approx 0.2$ at $h=1$, increases to about 0.4 in the domain
$0.25<h<0.5$, and then drops back to about 0.2 near $h\approx 0.15$.
In other words, the slope does not change much in each of these
broad domains: 
\begin{equation}
\frac{d\xi^*}{d(1/\sqrt{h})}
\approx 0.2 - 0.4\,,  
\end{equation}
or in common
units:
\begin{equation}
\frac{d\xi }{d(1/\sqrt{H})}\approx (0.2 -
0.4)\sqrt{\frac{\phi_0}{2\pi}}\,.  
\end{equation}
 Since the measured core size $\rho_c$ differs from our $\xi$ by a
constant shift, we may rewrite the last estimate as
\begin{equation}
\frac{d\rho_c}{d(1/\sqrt{H})}\,\approx (115 -230)\,\AA\sqrt{kOe}\,   
\end{equation}
in units employed in experiments.
Given our suggestion of ``universality", we expect the high-field 
slope $d\rho_c/d(1/\sqrt{H})$ for all materials to be in this
range.  For the set of materials discussed here, this is the
case:  we roughly estimate the slope as  220 for V$_3$Si, 190 for
NbSe$_2$, 240 for YNi$_2$B$_2$C, and 170 $\AA\sqrt{kOe}$ for
CeRu$_2$.   The ability of our approach to provide
  the slope values in a good agreement with the data indicates
that the model catches correctly the physics of the field dependence
of the core size. The slopes are relevant in particular, given an 
uncertain relationship between $\xi$ we calculate
and experimental $\rho_c$ (uncertain shifts $C$ in
$\xi^*(h)\approx \rho_c^*(h)+C(\lambda,T)$ shown in the middle panel
of Fig.\,\ref{fig2}). 

Still, a number of questions remains to be addressed.
Theoretically, it is not clear whether or not our   
clean limit results are compatible with the prediction of Kramer
and Pesh that the core size of an isolated vortex defined as 
$\rho_1=\Delta/(\partial\Delta/\partial r)_{r\to 0}$ goes to zero as
$T\to 0$.\cite{KP}  We note, however, that our results for $\rho_c$ 
are  meaningful only in large fields and for large GL parameters
$\kappa$, whereas these authors have considered an isolated vortex
in a material with $\kappa=0.9\,$. The same can be said with respect
to calculations of Ichioka {\it et al.} done for the d-wave symmetry
who find a shrinking core size in decreasing 
temperatures.\cite{Ichioka} 

Calculations of Miranovi\'c {\it et al.} of the low temperature
field dependence of the length
$\rho_1=\Delta_m/(\partial\Delta/\partial r)_{r\to 0}$ in the mixed
state ($\Delta_m$ is the order parameter maximum along the nearest
neighbor direction) show that for clean materials with $\lambda <1$
the length $\rho_1(H)$ goes through a minimum and increases
approaching $H_{c2}$. It also shows a much stronger field
dependence on the dirty side ($\lambda >1$) than our $\xi(H)$. A way
out of this difficulty, in our opinion, is to conclude that  
$\rho_1(H)$ is not proportional either to our $\xi(H)$  or  to
existing data on the core size $\rho_c(H)$ in clean materials for
which the minimum in $\rho_c(H)$ had not been recorded.\cite{Sonier1}

We have to mention that  our claim
that the field dependence of $\xi$ and $\rho_c$ should be suppressed
by impurity scattering and  disappear  altogether in the dirty
limit contradicts calculations of Golubov and Hartmann done within
the dirty limit Usadel formalism: they do find the $H$ dependence of
$\rho_c$ for NbSe$_2$.\cite{Golubov} Their conclusion have been
questioned in Ref.\,\onlinecite{NbSe2} on the grounds that the dirty
limit approximation does not hold for NbSe$_2$ crystals. Moreover, 
Nohara {\it et al.} in Ref.\,\onlinecite{YNiBC} report that the $H$
dependence of $\rho_c$  extracted from the field 
dependence of the specific heat coefficient $\gamma $ and well 
pronounced in pure NbSe$_2$, in fact disappears after doping
the crystal with Ta. The doping changes the impurity parameter from
0.19 to 1.25 so that the observation is consistent with our
conclusion. The data of this group on YNi$_2$B$_2$C and
Y(Ni$_{0.8}$Pt$_{0.2}$)B$_2$C are more convincing yet: the first 
crystal has the impurity parameter $\lambda\approx 0.4$ (i.e., it is
on the clean side) whereas the Pt-doped crystal is
on the dirty side with $\lambda\approx 2.4$; the $H$ dependence
of $\rho_c$ in the doped crystal is practically absent.  Still, the
controversy remains and could be resolved if the $\mu$SR data were  
 taken on a set of the same crystals with varying mean free
path; candidates for such a study could be, e.g, the Co-doped
LuNi$_2$B$_2$C crystals.\cite{Ames}

\acknowledgements

We are glad to thank J. Sonier for informative discussions and for
providing a table of the published data. Thoughtful remarks of
P. Miranovi\'{c} are greatly appreciated.  Ames Laboratory is
operated for the U. S. Department of Energy by Iowa State University
under Contract No. W-7405-Eng-82. This work is supported by the
Office of Basic Energy Sciences.

\appendix

\section{}

The sum (\ref{S2Dsum}) over $m,n$ can be replaced with the sum over
$m$ from 0 to $\mu=m+n$ and the sum over $\mu$ from 0 to $\infty$;
the former can be written as a hypergeometric function:
\begin{equation}
 S= \sum_{\mu=0}^{\infty} (2\mu-1)!!\,
\Big(-\frac{u}{2}\Big)^{\mu}\, _2F_1(-\mu,1-\sigma;1;2)\,. 
\label{**}
\end{equation}
We now use the integral representation
\begin{eqnarray}
_2F_1(a,b;c;z)&=&\frac{e^{-i\pi b}
z^{1-c}}{4\pi^2}\Gamma(c)\Gamma(1+b-c)\Gamma(1-b)\nonumber\\
&&\oint dt(t-z)^{c-b-1}t^{b-1}(1-t)^{-a}\,,
\label{Morse}
\end{eqnarray}
where the contour circles the branch points at $t=0$ and $t=z$ twice
in opposite directions; the representation holds everywhere except
points where the $\Gamma$-factors diverge.\cite{Morse} This yields:
\begin{eqnarray}
 S=\frac{e^{ i\pi (\sigma-1)}}{4\pi^2}
\Gamma(1-\sigma )\Gamma( \sigma)\oint
(t-2)^{\sigma-1}t^{-\sigma}\nonumber\\ 
\sum_{\mu=0}^{\infty} (2\mu-1)!!\,
\Big[-\frac{u}{2}(1-t)\Big]^{\mu} dt \,. 
\label{e9}
\end{eqnarray}
The sum here is transformed to an integral with the help of  
identity
\begin{equation}
\sum_{\mu=0}^{\infty}
(2\mu-1)!!\,(-x)^{\mu}=\frac{2}{\sqrt{\pi}}{\rm
Re}\int_0^{\infty}\frac{ds\, e^{-s^2}}{1+2s^2x}\,,
\label{e10}
\end{equation}
which is proven by formally expanding $1/(1+2s^2x)$ in powers of
$ 2s^2x$ and integrating over $s$.

 Then the contour integral emerges
of the form
\begin{equation}
J=\oint(t-2)^{\sigma-1}t^{-\sigma}\frac{dt}{1+s^2u(1-t)}\,,
\end{equation}
which can be transformed back to the hypergeometric form after the
substitution $t=v(1+s^2u)/s^2u=v \,z$:
\begin{equation}
J=\frac{4\pi^2
e^{i\pi(1-\sigma)}}{\Gamma(1-\sigma)\Gamma(\sigma)}\,
_2F_1(1,1-\sigma;1;2/z)\,.
\end{equation}
Further, $_2F_1(1,1-\sigma;1;2/z)=(1-2/z)^{\sigma-1}$ and we 
   obtain Eq.\,(\ref{S2D}) of the main text.

\section{}

One has to evaluate the integral at the RHS of Eq.(\ref{e19}); we
start with  the Fermi cylinder:
\begin{equation}
 \int_0^{ \omega_D}\frac{d\omega}{\omega}\,S =
\frac{2}{\sqrt{\pi}}{\rm Re}\int_0^{\infty}ds\,
e^{-s^2}\int_0^{\omega_D}\frac{d\omega}{\omega}\eta(\sigma,\omega)\,,
\label{A1}
\end{equation}
with
\begin{equation}
\eta=\frac{(1-s^2u)^{\sigma-1}}{(1+s^2u)^{\sigma}}\,,\quad
u=\frac{v^2q^2}{4\omega^2}\,.
\end{equation}
Substitution $x=\omega^2/\omega_D^2$ transforms the integral over
$\omega$ to
\begin{eqnarray}
&&J=\frac{1}{2}\int_0^1dx\frac{(x-y)^{\sigma-1}}{(x+y)^{\sigma}}
\nonumber\\
&&
=\frac{(-1)^{-\sigma}(1+y)^{1-\sigma}}{4y(\sigma-1)(2y)^{-\sigma}}\,
_2F_1\left(1-\sigma,1-\sigma;2-\sigma;\frac{1+y}{2y}\right)
\nonumber\\
&& -\frac{(-1)^{\sigma}}{4}\left[\psi\Big(\frac{1-\sigma}{2}\Big)-
\psi\Big(1-\frac{\sigma}{2}\Big)\right]\,,
\end{eqnarray}
where \begin{equation}
 y=\frac{s^2v^2q^2}{4\omega_D^2}\ll 1
\end{equation}
 because large values of $s$ are cut off by $e^{-s^2}$. 
Utilizing the reflection formulas for the Digamma function,\cite{Abr}
$\psi(1-z)=\psi(z)+\pi\cot(\pi z)$, the expression in square
parentheses is rewritten as
\begin{equation}
 \psi\Big(\frac{1+\sigma}{2}\Big)-
\psi\Big( \frac{\sigma}{2}\Big)-\frac{2\pi}{\sin(\pi
\sigma)} \,.
\end{equation}
We further use the asymptotic formula 15.3.13 of Ref. 
\onlinecite{Abr} for $_2F_1$ since  
$1/2y\gg 1$. Then, the real part of $J$ assumes the form:
\begin{eqnarray}
{\rm
Re}\,J=-\frac{\cos(\pi\sigma)}{4}\left[\psi\Big(\frac{1+\sigma}{2}\Big)-
\psi\Big( \frac{\sigma}{2}\Big) \right]\nonumber\\
-\frac{1}{2}\left[\ln 2+\gamma+\psi(\sigma)\right]
+\ln\frac{2\omega_D}{vq}-\ln s\,.
\label{B6}
\end{eqnarray}

  The integration over $s$ in Eq.\,(\ref{A1}) is now
straightforward: the $s$ independent part of Re$J$ enters the
result being unchanged because $(2/\sqrt{\pi})\int_0^{\infty}ds\,
e^{-s^2}=1$. Further: 
$(2/\sqrt{\pi})\int_0^{\infty}ds\, e^{-s^2}\ln s=-\gamma/2-\ln 2$.
 Collecting all terms in the self-consistency equation (\ref{e19}) we
obtain Eq.\,(\ref{***}).  As expected, the large parameter 
$\omega_D$ cancels out from the final result.

For the 3D situation, we have to replace in 2D Eq.\,(\ref{A1}) 
  $(2/\sqrt{\pi})\int_0^{\infty}ds\, e^{-s^2}{\rm
Re}\,J$ with 
$\sqrt{\pi}\int_0^{\infty}ds \,{\rm erfc}(s){\rm Re}\,J$. As in 2D, 
the $s$ independent part of Re$\,J$ enters the result being  
unchanged since
$\sqrt{\pi}\int_0^{\infty}ds \,{\rm erfc}(s)=1$, whereas
$\sqrt{\pi}\int_0^{\infty}ds \,{\rm erfc}(s)\ln s= -1-\gamma/2$. This
gives the 3D version of Eq.\,(\ref{***}).

\section{}

An explicitly real representation of the integral (\ref{S3D}) is
given in Ref.\,\onlinecite{K2} for the Fermi sphere. Here we provide
it for the 2D isotropic case of Eq.\,(\ref{S2D}). To this end, one 
separates the integration domain in two: $0<s<1/\sqrt{u}$ and
$ 1/\sqrt{u}<s<\infty$. In the first, the integration
variable is  changed to $y=s/\sqrt{u}$ whereas in the second to
$y=(s\sqrt{u})^{-1}$. Then we obtain:
\begin{eqnarray}
S(u,\sigma)&=&\frac{2}{\sqrt{\pi
u}}\int_0^1dy\frac{(1-y^2)^{\sigma-1}}{(1+y^2)^{\sigma}}
\\
&\times&\left[\exp\left(-\frac{y^2}{u}\right)-\cos(\pi\sigma)
\exp\left(-\frac{1}{y^2u}\right)\right]\,,\nonumber
\end{eqnarray}
the form easy to deal with numerically.


\begin{thebibliography}{99}
\bibitem{remark1} The function $\xi^*(h)$ for $h<0.15$ oscillates
with decreasing amplitude when  $h\to 0$ where
$\xi^*\to e^{\gamma/2}\approx 1.33$. Formally, 
the oscillations are due to the branch point of the integrands 
(\ref{S3D}) and (\ref{S2D}) which produces $\cos(\pi\sigma)$ in the
real part of the integral. The physical significance of this
zero-$T$ clean limit behavior is unclear.
\end{thebibliography}
\end{document}